\documentclass[pra,showpacs,twocolumn,superscriptaddress]{revtex4-1}
\usepackage{mathrsfs}
\usepackage{amsfonts}
\usepackage{amsmath}
\usepackage{txfonts}
\usepackage{amssymb}
\usepackage{graphicx}
\usepackage{bm}
\usepackage{color}
\usepackage[normalem]{ulem}

\newcommand{\ket}[1]{|#1\rangle}
\newcommand{\bra}[1]{\langle #1|}

\begin{document}
\title{$d$-pod realization of nonadiabatic holonomic quantum computation}
\author{Oskar Axelsson}
\affiliation{Department of Physics and Astronomy, Uppsala University,
Box 524, Se-751 20 Uppsala, Sweden}
\author{Claes F\"alth}
\affiliation{Department of Physics and Astronomy, Uppsala University,
Box 524, Se-751 20 Uppsala, Sweden}
\author{Elias Henriksson Lindberg}
\affiliation{Department of Physics and Astronomy, Uppsala University,
Box 524, Se-751 20 Uppsala, Sweden}
\author{Erik Sj\"oqvist}
\email{erik.sjoqvist@physics.uu.se}
\affiliation{Department of Physics and Astronomy, Uppsala University,
Box 524, Se-751 20 Uppsala, Sweden}
\date{\today}
\begin{abstract}
Holonomic quantum computation (HQC) realizes quantum gates through non-Abelian 
geometric phases, providing an experimentally accessible approach to quantum control. 
While the nonadiabatic HQC framework has been extensively developed for three-level 
$\Lambda$ systems encoding qubits, its systematic extension to higher-dimensional 
qudits remains largely unexplored. In this work, we generalize nonadiabatic HQC to a 
$d$-pod configuration, where a single excited state is coupled to $d$ ground states, 
the latter forming the computational subspace. This scheme enables universal holonomic 
single- and two-qudit gates using only optical or microwave pulses on 
trapped atoms or ions, offering an efficient route to implement a discrete universal gate 
set with minimal pulse coordination. As an explicit example, we analyze in detail the 
qutrit ($d=3$) case, demonstrating compact realizations of single- and two-qutrit 
holonomic gates, each gate requiring at most two loops in the Grassmannian generated 
by at most three pulses. 
\end{abstract}
\maketitle
\date{\today}

\section{Introduction}
Holonomic quantum computation (HQC), originally proposed for adiabatic systems 
\cite{zanardi99} and later extended to the nonadiabatic regime \cite{sjoqvist12}, exploits 
non-Abelian (matrix-valued) geometric phases \cite{wilczek84,anandan88} to implement 
quantum gates. This idea has developed into a well-established approach to qubit-based 
quantum computation 
\cite{duan01,wu05,xu12,zhang14,xu14,zhang15,xu15,sjoqvist16,herterich16,xue17,xu18a,ramberg19,liu19} 
with multiple experimental demonstrations in recent years 
\cite{abdumalikov13,toyoda13,feng13,arroyo14,zu14,zhou17,sekiguchi17,danilin18,nagata18,leroux18,xu18b,yan19,ai20,zhao21a,lu25}. 
Nonadiabatic holonomic quantum computation enables fast geometric quantum gates 
with intrinsic robustness to certain control errors 
\cite{sjoqvist12,johansson12,sun16,xu17a,xu17b,zhao21b,alves22}, making it a promising 
ingredient for fault-tolerant quantum computing architectures when combined with quantum 
error correction. For a comprehensive review, see Ref.~\cite{zhang23}. 

The standard approach to nonadiabatic HQC \cite{sjoqvist12} employs a $\Lambda$-type 
configuration, where a single excited state is coupled to two ground state levels that span 
the qubit state space. The inclusion of the excited state allows the qubit space to be embedded 
as a proper subspace of the full state space; a necessary requirement for realizing nontrivial 
holonomies. Geometrically, these holonomies correspond to loops in the Grassmannian 
$\mathcal{G}(3;2)$, the manifold of two-dimensional subspaces of a three-dimensional 
Hilbert space.

While most quantum computing architectures are based on qubits, higher-dimensional 
systems ($d$-level qudits with $d\ge3$) offer the potential for exponentially increased 
information capacity as the number of subsystems grows. They also enable efficient circuit 
implementations \cite{lanyon09} and possess advantageous properties for fault-tolerant 
quantum computation \cite{campbell12,campbell14}, potentially easing the demands of 
reaching the threshold for effective quantum error correction 
\cite{andrist15,michael16,muralidharan17}. Experimentally, qudits have been realized 
on various physical platforms \cite{kues17,moro19,soltamov19,hrmo23,meng24}.

In this work, we extend nonadiabatic holonomic control from qubits to qudits by generalizing 
the standard $\Lambda$-scheme to a $d$-pod configuration \cite{rosseaux13}, in which 
$d$ computational states are coupled to a single auxiliary excited state. Although nonadiabatic 
holonomic qudit schemes employing multiple auxiliary excited states have been investigated 
\cite{azimi14a,azimi14b,xu21,andre22}, and multilevel holonomic control has been studied 
in the adiabatic regime \cite{recati02,karle03,shkolnikov20,cuadra22}, a systematic treatment 
of nonadiabatic holonomies for the single-excited-state $d$-pod configuration remains lacking. 
Our approach, which can be viewed as a nonadiabatic extension of Ref.~\cite{recati02}, 
provides a compact characterization of the accessible holonomies on the Grassmannian 
$\mathcal{G}(d+1;d)$ and enables resource-efficient universal qudit computation 
using optical or microwave driving fields. As a measure of efficiency, we analyze the number 
of driving pulses required to implement a universal discrete set of single-qudit holonomic 
gates together with an entangling two-qudit gate.

The proposed $d$-pod setting builds upon the multi-pulse single-loop method introduced in 
Ref.~\cite{herterich16}, thereby minimizing the path length in the Grassmannian. Although 
the construction applies to arbitrary $d$, as outlined in the next section, we focus the detailed 
analysis on the qutrit ($d=3$) case in Sec.~\ref{sec:qutrit}. We conclude in Sec.~\ref{sec:conclusion}.   

\section{Holonomic qudit gates in the $d$-pod setting}
\subsection{The $d$-pod}
Qubits are typically realized in quantum systems with higher-dimensional Hilbert spaces 
by retaining only two levels, thereby defining a two-level computational subspace. The 
qudit generalization of this approach involves more than two levels within the same physical 
system to form a higher-dimensional computational subspace. A register of $n$ qudits can 
exist in a superposition of $d^n$ computational basis states, meaning that the exponential 
information capacity grows faster with the dimensionality of the primitive information carriers. 
This makes qudits attractive for high-density encoding of quantum information.

\begin{figure}[h!]
\centering
\includegraphics[width=0.45\textwidth]{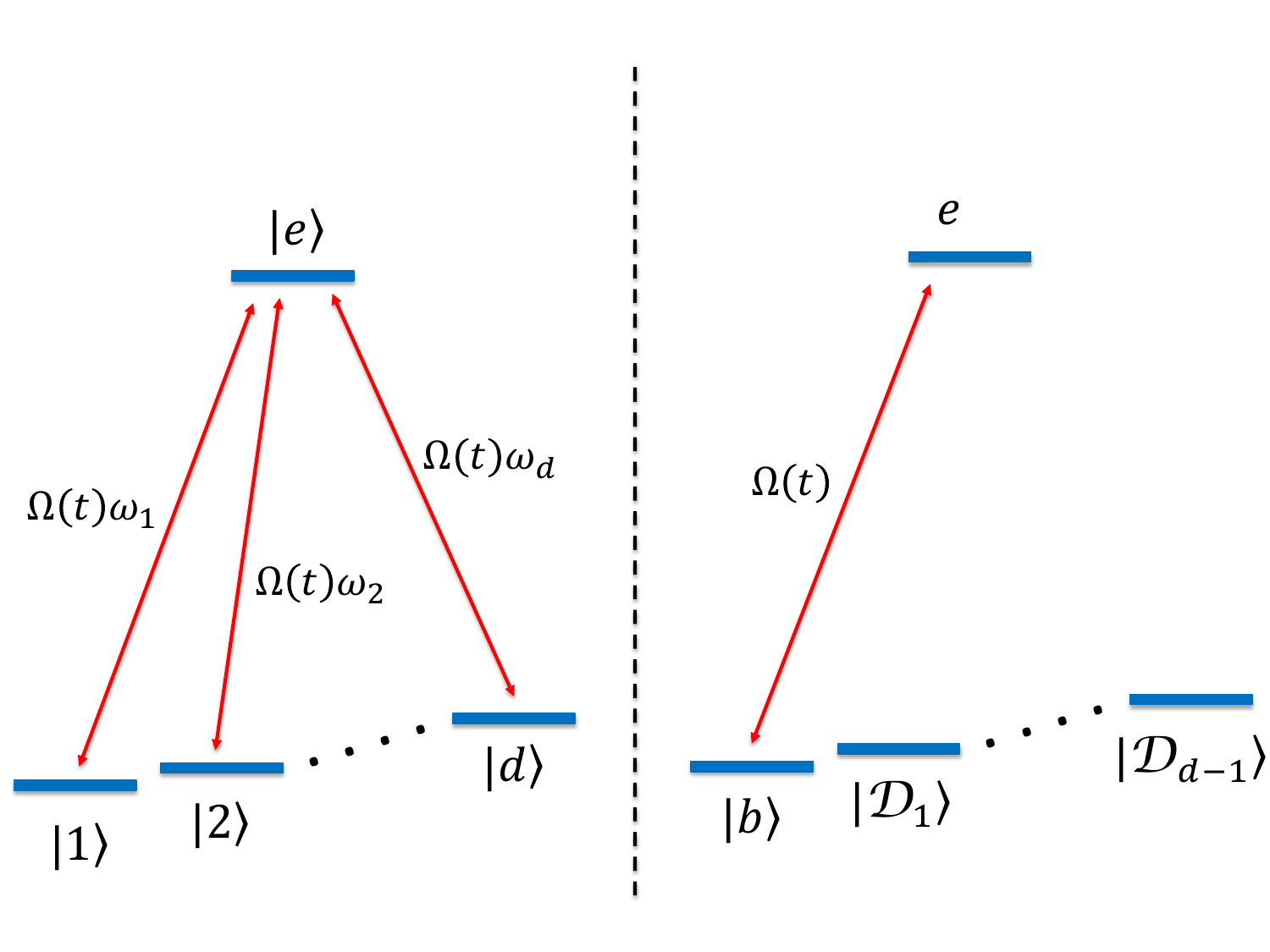}
\caption{Setup for a qudit system implemented using the $d$-pod configuration. Left panel: 
Each computational basis state $\ket{1},\dots,\ket{d}$ is coupled to an excited state $\ket{e}$. 
Right panel: The same system expressed in the dark-bright Morris-Shore \cite{morris83} basis 
$\{ \ket{b}, \ket{\mathcal{D}_1}, \ldots, \ket{\mathcal{D}_{d-1}} \}$. Note that only the bright 
state $\ket{b}$ is coupled to $\ket{e}$.}
\label{fig:d_pod_1}
\end{figure}

We consider the $d$-pod setting, shown in Fig.~\ref{fig:d_pod_1}, which realizes a single 
qudit in the $d$ ground state levels $1, \ldots, d$ coupled to a single excited state 
$e$ of, e.g., a trapped atom or ion. These couplings are assumed to be mediated by a 
set of simultaneously coordinated optical or microwave field pulses, all having the same 
temporal envelope shape, as represented by the single real-valued function $\Omega (t)$. 
The fields differ in their time-independent amplitudes and phases, as captured by the 
complex-valued parameters $\omega_1,\ldots,\omega_d$. The Hamiltonian that describes 
the system reads ($\hbar = 1$ from now on)
\begin{eqnarray}
H(t) & = &  
\sum_{k=1}^{d} \Omega (t) \big( \omega_k \ket{e} \bra{k} + \omega_k^{\ast} \ket{k} \bra{e} \big) 
\nonumber \\ 
 & \equiv & \Omega (t) \big( \ket{e} \bra{b} + 
\ket{b} \bra{e} \big) 
\label{eq:H}
\end{eqnarray}
with the bright state 
\begin{eqnarray}
\ket{b} = \sum_{k=1}^{d} \omega_k^{\ast} \ket{k}.
\end{eqnarray}
The $d-1$ dark states 
\begin{eqnarray}
\ket{\mathcal{D}_j} = \sum_{k=1}^{d} c_k^{(j)} \ket{k}    
\end{eqnarray}
form an orthonormal set satisfying 
\begin{eqnarray}
\sum_{k=1}^{d} c_k^{(j)} \omega_k = 0
\end{eqnarray}
with $j=1,\ldots,d-1$, and are decoupled from the excited state and hence from the dynamics. 
The $d$-dimensional ground state subspace is spanned by the Morris-Shore basis states 
\cite{morris83} $\ket{b}, \ket{\mathcal{D}_1}, \ldots, \ket{\mathcal{D}_{d-1}}$. The Hamiltonian 
follows by assuming monochromatic fields that couple the excited state to each ground state 
level of the atom or ion and by using the rotating wave approximation in a frame co-rotating 
with the fields. 

\subsection{Holonomic qudit gates in the $d$-pod}
A qudit gate in the above $d$-pod system is a unitary operator acting on the ground state 
manifold ${\rm Span} \{ \ket{1}, \ldots, \ket{d} \}$, forming the computational subspace 
$\mathcal{M}$. Explicitly, such a gate is realized if there exists a run time $\tau$ so that
\begin{eqnarray}
\mathbb{U}_{kl} & = & \bra{k} \mathcal{T} e^{-i \int_0^{\tau} H(s) ds} \ket{l} 
\end{eqnarray}
with $\mathcal{T}$ time ordering and $k,l=1,\ldots,d$, is unitary. A universal qudit gate set 
can be obtained from suitable compositions of the elementary holonomies generated by 
Eq.~(\ref{eq:H}), together with any entangling two-qudit gate \cite{sawicki17,wang20}. 

The conditions to assure the purely holonomic nature of the gate induced by the Hamiltonian 
in Eq.~(\ref{eq:H}) are: (i) the time dependent envelope $\Omega (t)$ should have a finite 
duration $\tau$ (the `run time' of the gate) and be the same for all fields; (ii) the laser 
parameters $\omega_k$ should be time independent over the duration of each pulse; 
(iii) the optical fields with frequencies $\nu_k$ should be resonant to the respective Bohr 
frequencies $f_{ek}$, i.e., $\nu_k=f_{ek}$. 

To demonstrate the sufficiency of the above conditions, we note that they imply 
$[H(t),U(t,0)] = 0$, $t \in [0,\tau]$, with $U(t,0) = \mathcal{T} e^{-i \int_0^t H(s) ds} = 
e^{-i \int_0^t \Omega(s) ds (\ket{e}\bra{b}+\ket{b}\bra{e})}$ the time evolution operator, 
from which follows that 
\begin{eqnarray}
\mathbb{K}_{kl} (t) & \equiv & \bra{k} U^{\dagger} (t,0) H(t) U(t,0) \ket{l} = 0, 
\nonumber \\ 
 & & \forall \ k,l=1,\ldots,d.
 \label{eq:pt}
\end{eqnarray}
Thus, the non-Abelian dynamical phase \cite{anandan88} $\mathcal{T} e^{-i \int_0^{\tau} 
\mathbb{K} (t)dt}$ is trivial on $\mathcal{M}$. 

Geometrically, the computational subspace $\mathcal{M}$ evolves as a point in the 
Grassmannian manifold $\mathcal{G}(d+1;d)$ of $d$-dimensional subspaces of a 
$(d+1)$-dimensional Hilbert space. The bright state determines the direction in which 
$\mathcal{M}$ is displaced away from its initial position, while the dark states remain 
fixed. A cyclic evolution of $\mathcal{M}$ thus corresponds to a closed loop in 
$\mathcal{G}(d+1;d)$, with the resulting gate given by the associated non-Abelian 
holonomy. The evolution becomes cyclic when the duration $\tau$ is chosen such that
\begin{eqnarray}
\int_0^{\tau} \Omega(t)dt = \pi ,
\end{eqnarray}
since, for this choice of $\tau$, the evolution operator $U(\tau,0)$ leaves $\mathcal{M}$ invariant:
\begin{eqnarray}
U(\tau,0) P_d U^{\dagger}(\tau,0)=P_d,
\end{eqnarray}
where
\begin{eqnarray}
P_d=\ket{1}\bra{1}+\ldots+\ket{d}\bra{d}
\end{eqnarray}
is the projector onto $\mathcal{M}$. As the dynamical contribution vanishes identically 
throughout the evolution, the resulting transformation is purely holonomic.

Consider a cyclic evolution generated by two pulse segments of equal pulse area, 
characterized by laser parameters $\{\omega_k\}$ and $\{e^{i\eta}\omega_k\}$, respectively. 
The induced holonomy on the computational subspace $\mathcal{M}$ is 
\begin{equation}
\mathbb{U} =
P_d-(1+e^{-i\eta})|b\rangle\langle b|,
\end{equation}
where $|b\rangle$ is the bright state and $P_d$ is the projector onto $\mathcal{M}$.

The operator $\mathbb{U}$ may be viewed as a complex Householder transformation 
acting on the computational subspace \cite{ivanov06}. Products of such transformations generate a 
rich class of unitary operations and form the basis of the gate constructions developed 
below. For $\eta=0$ it reduces to a reflection about the hyperplane orthogonal to the 
bright state,
\begin{equation}
\mathbb{U}=P_d-2|b\rangle\langle b|,
\end{equation}
while arbitrary $\eta$ introduces an additional geometric phase. This observation provides 
a useful interpretation of the elementary holonomic gates generated in the $d$-pod architecture.

\subsection{Entangling holonomic two-qudit gates}
\label{sec:2qudit}
To achieve all-geometric universal quantum computation, one has to realize at least one 
gate that can entangle a pair of qudits. By adapting Ref.~\cite{andre22} to the $d$-pod, 
one can implement a conditional two-qudit gate of the form
\begin{eqnarray}
\tilde{\mathbb{U}} = \begin{pmatrix}
\mathbb{I}_{d^2-d} & \\ 
 & \mathbb{V}_d
\end{pmatrix} . 
\end{eqnarray}
The physical setting consists of two trapped ions whose internal levels encode qudits and 
which interact through a common vibrational mode of the trap. One ion acts as a control 
qudit while the second ion serves as the target. By applying suitably detuned laser fields 
in a S{\o}rensen-M{\o}lmer configuration \cite{sorensen99}, the vibrational degree of freedom 
can be adiabatically eliminated, yielding an effective interaction between the internal states 
of the ions. As shown in Ref.~\cite{andre22}, the resulting dynamics can be described by 
the effective Hamiltonian
\begin{eqnarray}
H_{\rm eff} (t) = \ket{d}\bra{e}\otimes H(t) + {\rm H.c.},
\end{eqnarray}
where $H(t)$ is the single-qudit $d$-pod Hamiltonian in Eq.~\eqref{eq:H} acting on the target ion.

The key feature of $H_{\rm eff} (t)$ is that the $d$-pod dynamics of the target qudit is activated 
only when the control ion occupies the state $\ket{d}$. Thus, the computational basis states
\begin{eqnarray}
\ket{m}\otimes\ket{k},
\quad
m\neq d,
\end{eqnarray}
remain unaffected, while the subspace
\begin{eqnarray}
\ket{d}\otimes {\rm Span} \{ \ket{1},\ldots,\ket{d} \}
\end{eqnarray}
undergoes precisely the same holonomic evolution as a single qudit in the $d$-pod 
configuration. Choosing the pulse sequence to generate a nontrivial single-qudit holonomy 
$\mathbb{V}_d$, one obtains the controlled operation $\tilde{\mathbb{U}}$ above. When 
$\mathbb{V}_d\neq e^{i\phi} \mathbb{I}_d$, the gate is entangling and therefore cannot 
be decomposed into independent single-qudit transformations.

The holonomic nature of the gate follows from the same conditions as in the single-qudit case: 
the evolution is cyclic within the relevant computational subspace and satisfies the 
parallel-transport condition Eq.~\eqref{eq:pt}, ensuring the absence of dynamical 
contributions. Consequently, the acquired transformation is purely holonomic. Together 
with a suitable set of single-qudit holonomic gates generated in the $d$-pod setting, the 
conditional gate $\tilde{\mathbb{U}}$ provides a universal set of holonomic gates for qudit 
computation \cite{sawicki17,wang20}.

\section{Example: Qutrit holonomic gates in the tripod}
\label{sec:qutrit}
Here, we demonstrate how the above general scheme can be used to implement a universal 
set of qutrit ($d=3$) gates. In the case of one-qutrit gates, we use the parameterization 
\begin{eqnarray}
\omega_1 & = & \sin \theta \cos \varphi , 
\nonumber \\ 
\omega_2 & = & e^{i\alpha} \sin \theta \sin \varphi , 
\nonumber \\ 
\omega_3 & = & e^{i\beta} \cos \theta,
\end{eqnarray}
where $\theta \in [0,\pi]$ and $\varphi,\alpha,\beta \in [0, 2\pi)$. This results in the holonomy 
\begin{widetext}
\begin{eqnarray}
\mathbb{U} \equiv 
\mathbb{U} (\theta,\varphi,\alpha,\beta,\eta) = 
\begin{pmatrix}
1-f(\eta) \sin^2 \theta \cos^2 \varphi & -f(\eta) e^{i\alpha} \sin^2 \theta \sin \varphi \cos \varphi & 
-f(\eta) e^{i\beta} \sin \theta \cos \theta \cos \varphi \\
-f(\eta) e^{-i\alpha} \sin^2 \theta \sin \varphi \cos \varphi & 1-f(\eta)\sin^2 \theta \sin^2 \varphi & 
-f(\eta) e^{i(\beta -\alpha)} \sin \theta \cos \theta \sin \varphi \\
-f(\eta) e^{-i\beta} \sin \theta \cos \theta \cos \varphi & 
-f(\eta) e^{-i(\beta -\alpha)} \sin \theta \cos \theta \sin \varphi & 1-f(\eta) \cos^2 \theta 
\end{pmatrix}
\label{eq:gengate}
\end{eqnarray}
\end{widetext}
with $f(\eta) = 1+e^{-i\eta}$, associated with a single loop. Note that 
$\mathbb{U} (\theta,\varphi,\alpha,\beta,\pi) = \mathbb{I}$, $\mathbb{I}$ being the 
$3\times 3$ identity matrix, as the loop corresponding to $\eta=\pi$ encloses no area. 
Important holonomic qutrit gates are diagonal phase shift gates, Pauli gates,  Hadamard, 
and an entangling two-qutrit gate, as will be discussed in the following three subsections. 
In particular, a universal set can be taken as the qutrit Hadamard gate, the non-Clifford 
$\mathbb{T}_3$ gate, and an entangling two-qutrit gate. 

\subsection{Pauli and $\mathbb{T}_3$ gates}
We first consider holonomic implementation of diagonal phase shift gates in the $d$-pod setting.
For this class of gates, we may note that the only diagonal unitaries that can be achieved with 
a single-loop must have $\theta = 0$ and therefore be of the form 
\begin{eqnarray}
\mathbb{U} (0,\varphi,\alpha,\beta,\eta) = 
\begin{pmatrix}
1 & 0 & 0 \\
0 & 1 & 0 \\
0 & 0 & 1-f(\eta) 
\end{pmatrix} = 
\begin{pmatrix}
1 & 0 & 0 \\
0 & 1 & 0 \\
0 & 0 & -e^{-i\eta} 
\end{pmatrix} .  
\end{eqnarray}
This acts non-trivially only on a `qubit' subspace ${\rm Span} \{ \ket{2},\ket{3} \}$, say, 
thereby making them useless for qutrit universality. We therefore need at least two loops 
to achieve genuine qudit phase shift gates. To this end, we use   
\begin{eqnarray}
\mathbb{U} \left(\frac{\pi}{4},\frac{\pi}{2},\alpha,\beta,0 \right) = 
\begin{pmatrix}
1 & 0 & 0 \\
0 & 0 & - e^{i(\beta -\alpha)} \\
0 & - e^{-i(\beta -\alpha)} & 0
\end{pmatrix} , 
\end{eqnarray}
which implies 
\begin{eqnarray}
& & \mathbb{U} \left( \frac{\pi}{4}, \frac{\pi}{2}, \alpha', \beta',0\right) \mathbb{U} 
\left(\frac{\pi}{4}, \frac{\pi}{2}, \alpha,\beta,0 \right) 
\nonumber \\ 
 & = &
\begin{pmatrix}
1 & 0 & 0 \\
0 & e^{i(\beta -\alpha - \beta' +\alpha' )} & 0 \\
0 & 0 & - e^{-i(\beta -\alpha - \beta' +\alpha' )} 
\end{pmatrix} . 
\end{eqnarray}
This implements the Pauli gate  
\begin{eqnarray}
\mathbb{Z}_3 = \begin{pmatrix}
1 & 0 & 0 \\
0 & e^{i\frac{2\pi}{3}} & 0 \\
0 & 0 & e^{i\frac{4\pi}{3}} \\
\end{pmatrix}, 
\label{eq:Z-gate}
\end{eqnarray}
and the non-Clifford  
\begin{eqnarray}
\mathbb{T}_3 = \begin{pmatrix}
1 & 0 & 0 \\
0 & e^{i\frac{2\pi}{9}} & 0 \\
0 & 0 & e^{-i\frac{2\pi}{9}} \\
\end{pmatrix}
\label{eq:T-gate}
\end{eqnarray}
by choosing $\beta -\alpha - \beta' +\alpha' = \frac{2\pi}{3}$ and 
$\beta -\alpha - \beta' +\alpha' = \frac{2\pi}{9}$, respectively. As $\eta = 0$ realizes 
a single pulse per loop, these gates are two-pulse implementations each defining 
two loops in the Grassmannian. 

Next, we complete the set of Pauli gates by implementing a holonomic version of the 
off-diagonal Pauli-$X$ (shift) gate: 
\begin{equation}
\mathbb{X}_3 =
\begin{pmatrix}
 0 & 0 & 1 \\
 1 & 0 & 0 \\
 0 & 1 & 0 \\
\end{pmatrix}. 
\label{eq:X-gate}
\end{equation}
This gate may be realized by noting that the parameter choices 
$(\theta,\varphi,\alpha,\beta,\eta)=(\frac{\pi}{2},\frac{\pi}{4},\pi,0,0)$ and 
$(\theta,\varphi,\alpha,\beta,\eta)=(\frac{\pi}{4},\frac{\pi}{2},0,\pi,0)$ correspond to gates 
that flip the states $\ket{0} \leftrightarrow \ket{1}$ and $\ket{1} \leftrightarrow \ket{2}$, 
respectively. Explicitly, one finds 
\begin{eqnarray}
\mathbb{U} \left( \frac{\pi}{2},\frac{\pi}{4},\pi,0,0 \right) = \begin{pmatrix}
0 & 1 & 0 \\ 
1 & 0 & 0 \\ 
0 & 0 & 1  
\end{pmatrix}
\end{eqnarray}
and 
\begin{eqnarray}
\mathbb{U} \left( \frac{\pi}{4},\frac{\pi}{2},0,\pi,0 \right) = \begin{pmatrix}
1 & 0 & 0 \\ 
0 & 0 & 1 \\ 
0 & 1 & 0  
\end{pmatrix} . 
\end{eqnarray}
Thus, with two loops of the computational subspace 
one obtains 
\begin{eqnarray}
\mathbb{X}_3 = \mathbb{U}\left(\frac{\pi}{2}, \frac{\pi}{4}, \pi, 0,0\right)\mathbb{U} 
\left(\frac{\pi}{4}, \frac{\pi}{2}, 0, \pi,0\right) . 
\label{eq:X_realization}
\end{eqnarray}
Stated differently, these two loops implement the consecutive permutations 
$(0,1,2) \rightarrow (0,2,1)\rightarrow (2,0,1)$, which indeed is the effect of 
$\mathbb{X}_3$. Similarly, $\mathbb{X}_3^{\rm T}$ is implemented the two loops 
in reverse order, resulting in the consecutive permutations $(0,1,2) \rightarrow (1,0,2) 
\rightarrow (1,2,0)$. Again, as each loop is implemented with $\eta = 0$, only one 
pulse per loop is needed. 

\subsection{Hadamard gate}
As seen, the $\mathbb{X}_3$, $\mathbb{Z}_3$, and $\mathbb{T}_3$ gates are relatively 
simple to find by fixing the parameters in the correct positions, by trying different angles 
for $\theta$ and $\varphi$, and thereafter fixing the exponents using $\alpha$ and $\beta$. 
Further, they all can be implemented with one pulse per loop, i.e., by keeping $\eta = 0$ 
throughout. In contrast, the Hadamard gate 
\begin{equation}
\mathbb{H}_3 =
\frac{1}{\sqrt{3}}
\begin{pmatrix}
 1 & 1 & 1 \\
 1 & e^{i\frac{2\pi}{3}} & e^{i\frac{4\pi}{3}} \\
 1 & e^{i\frac{4\pi}{3}} & e^{i\frac{2\pi}{3}} \\
\end{pmatrix}, 
\label{eq:Hadamard-gate}
\end{equation}
is considerably more challenging to realize analytically. Unlike the gates considered 
above, inspection of Eq.~(\ref{eq:gengate}) does not immediately reveal a parameter 
choice yielding the desired transformation.

While it is not even apparent how many pulses are needed here, we may exclude any 
single loop implementation, as such a loop must be generated by a pulse pair with 
$\eta=\frac{\pi}{2}$ due to the constraint $\det \mathbb{H}_3 = -i$. This leads to a direct 
contradiction as the upper left element of $\mathbb{U}(\theta,\varphi,\alpha,\beta,\eta)$ 
with $\eta = \frac{\pi}{2}$ would be a complex number. On the other hand, by combining 
two loops and by numerically minimizing the Frobenius norm 
\begin{eqnarray}
\| \mathbb{A} \|_F = \sqrt{ \operatorname{Tr}(\mathbb{A}^{\dagger} \mathbb{A})}
\end{eqnarray}
with 
\begin{eqnarray}
\mathbb{A} = \mathbb{U} (\theta',\varphi',\alpha',\beta',\eta') 
\mathbb{U} (\theta,\varphi,\alpha,\beta,\eta) - \mathbb{H}_3 , 
\end{eqnarray}
one finds excellent agreement. The optimization was performed by minimizing 
$\|\mathbb{A}\|_F$ over the ten free parameters associated with the two loops. 
The resulting minimum satisfies
\begin{equation}
\| \mathbb{A}\|_F \approx 10^{-4},
\end{equation}
demonstrating that the obtained pulse sequence reproduces the target gate to high 
numerical accuracy. Explicitly, such a minimization yields 
\begin{eqnarray}
(\theta,\varphi,\alpha,\beta,\eta) =(0.8918,2.2028,0.0006,-0.0006,0)  
\end{eqnarray}
for the first pulse, and, interestingly, numerical values that are remarkably close to the 
analytical expressions
\begin{eqnarray}
\left(\theta',\varphi',\alpha',\beta',\eta'\right)  = 
\left( \frac{\pi}{4}, \frac{\pi}{2} ,-\frac{\pi}{5},\frac{4\pi}{5},\frac{\pi}{2} \right)  
\end{eqnarray}
for the second pulse. By inserting these parameter values into the general expression 
in Eq.~\eqref{eq:gengate}, one finds 
\begin{widetext}
\begin{eqnarray}
\mathbb{U} \left( \frac{\pi}{4}, \frac{\pi}{2} ,-\frac{\pi}{5},\frac{4\pi}{5},\frac{\pi}{2} \right)  
\ \mathbb{U} \left( 0.8918,2.2028,0.0006,-0.0006,0 \right) 
= \begin{pmatrix}
0.5774 & 0.5774 & 0.5774 \\ 
0.5774 & -0.2887 + 0.5000 i & -0.2887 - 0.5000 i \\ 
0.5774 & -0.2887 - 0.5000 i & -0.2887 + 0.5000 i
\end{pmatrix}
\end{eqnarray}
\end{widetext}
which indeed is an accurate approximation of  $\mathbb{H}_3$ in Eq.~\eqref{eq:Hadamard-gate}. 
While the second loop is associated with a non-zero $\eta$ and therefore requires two pulses, 
the first loop can be implemented with a single pulse. We therefore conclude that $\mathbb{H}_3$ 
admits a three-pulse implementation. Since a single-loop realization is excluded by the determinant 
constraint discussed above, and the present construction uses the smallest number of loops 
for which a solution was found, three pulses constitute the shortest implementation within the 
present framework.

We note that, in the dark-path holonomic qudit scheme developed in Ref.~\cite{andre22}, the 
realization of $\mathbb{H}_3$ required numerical optimization, whereas analytical expressions 
for the control parameters were obtained for the other gates, namely $\mathbb{X}_3$, 
$\mathbb{Z}_3$, and $\mathbb{T}_3$. This may indicate that the holonomic implementation 
of the qutrit Hadamard gate is exceptional in terms of pulse parametrization, as it appears to 
be less amenable to a fully analytical treatment than the other elementary qutrit gates considered.

\subsection{Two-qutrit gate}
The conditional two-qudit construction described in Sec.~\ref{sec:2qudit} applies directly to 
the qutrit case ($d=3$). By choosing the target-space holonomy $\mathbb{V}_3$ to be any 
nontrivial single-qutrit gate generated in the tripod configuration, one obtains the controlled 
operation
\begin{equation}
\tilde{\mathbb{U}}_3=
\begin{pmatrix}
\mathbb{I}_6 & \\
& \mathbb{V}_3
\end{pmatrix},
\end{equation}
written in a basis where the final three-dimensional block corresponds to the control qutrit 
occupying the state $\ket{3}$. Thus, the target qutrit undergoes the holonomic transformation 
$\mathbb{V}_3$ only when the control qutrit is in the state $\ket{3}$, while all other computational 
basis states remain unchanged. A particular choice is to take $\mathbb{V}_3=\mathbb{Z}_3$. 
This gate can be realized by applyinging sequentially the two conditional Hamiltonians 
\begin{eqnarray}
H_{\rm eff}^{(1)}(t)
& = &
\frac{\Omega_1(t)}{\sqrt{2}}
\ket{3} \bra{e}
\otimes
\Big[ \ket{e} \left(
\bra{2}+\bra{3} \right)
\nonumber \\ 
 & & + \left( \ket{2}+\ket{3} \right)
\bra{e} \Big] + {\rm H.c.},
\nonumber \\ 
H_{\rm eff}^{(2)}(t)
& = & \frac{\Omega_2(t)}{\sqrt{2}}
\ket{3}\bra{e} \otimes
\Big[ \ket{e} \Big(
e^{i2\pi/3}\bra{2}+\bra{3}
\Big)
\nonumber \\ 
 & & + \Big( e^{-i2\pi/3} 
\ket{2}+\ket{3} \Big) \bra{e}
\Big] + {\rm H.c.} 
\label{eq:Heff_Z3_2}
\end{eqnarray}
The first pulse couples the target excited state $\ket{e}$ to the bright state
\begin{eqnarray}
\ket{b_1} = \frac{1}{\sqrt{2}}
\big( \ket{2}+\ket{3} \big),
\end{eqnarray}
whereas the second pulse couples $\ket{e}$ to
\begin{eqnarray}
\ket{b_2} = \frac{1}{\sqrt{2}}
\big( e^{-i2\pi/3}\ket{2}+\ket{3} \big).
\end{eqnarray}
In both cases, the interaction is conditioned on the control qutrit occupying the state $\ket{3}$, 
while all states with the control qutrit in $\ket{1}$ or $\ket{2}$ remain unaffected. Since $\eta=0$ 
for both loops, each pulse generates a single-loop holonomy of the form
\begin{eqnarray}
\mathbb{U}_j = \begin{pmatrix}
\mathbb{I}_6 & \\
&
\mathbb{I}_3-2\ket{b_j}\bra{b_j}
\end{pmatrix} , 
\quad
j=1,2.
\end{eqnarray}
The two successive holonomies give
\begin{eqnarray}
\mathbb{U}_2\mathbb{U}_1 = 
\begin{pmatrix}
\mathbb{I}_6 & \\
&
\mathbb{Z}_3
\end{pmatrix} , 
\end{eqnarray}
Thus, the target qutrit undergoes the $\mathbb{Z}_3$ transformation only when the 
control qutrit is in $\ket{3}$, while all other computational basis states remain unchanged. 
Since $\mathbb{Z}_3$ is not proportional to the identity, $\mathbb{U}_2\mathbb{U}_1$ 
is an entangling gate. The construction therefore provides a two-pulse, fully holonomic 
realization of the desired controlled-$\mathbb{Z}_3$ gate. Together with the holonomic 
single-qutrit gates $\mathbb{X}_3$, $\mathbb{Z}_3$, $\mathbb{T}_3$, and $\mathbb{H}_3$ 
constructed above, this provides a universal set of holonomic gates for qutrit quantum computation.

\section{Conclusions}
\label{sec:conclusion}
We have extended nonadiabatic holonomic quantum computation from the standard 
three-level $\Lambda$ configuration to a $d$-pod architecture consisting of $d$ computational 
states coupled to a single auxiliary excited state. The resulting scheme implements purely 
holonomic single- and two-qudit gates as non-Abelian holonomies associated with loops in 
the Grassmannian $\mathcal{G}(d+1;d)$. By using the single-loop multi-pulse construction of 
Ref.~\cite{herterich16}, we derived a compact expression for the elementary holonomies 
accessible in the $d$-pod configuration and showed how these holonomies may be combined 
to construct universal qudit gate sets. This scheme can be implemented using internal states 
of trapped atoms or ions. 

To illustrate the method, we analyzed the qutrit case in detail. Explicit pulse sequences were 
derived for the generalized Pauli gates $\mathbb{X}_3$ and $\mathbb{Z}_3$, as well as for 
the non-Clifford gate $\mathbb{T}_3$. These gates were shown to admit optimal implementations 
requiring only two pulses. The qutrit Hadamard gate is more demanding, requiring two loops 
and a minimum of three pulses, highlighting a qualitative distinction between the Hadamard gate 
and the other generators of a universal qutrit gate set. Taken together, these results identify 
the $d$-pod architecture as a conceptually simple and experimentally resource-efficient 
platform for nonadiabatic holonomic control of higher-dimensional quantum systems. By 
using only a single auxiliary excited state, the scheme combines minimal hardware overhead 
with a transparent geometric interpretation, making it a promising candidate for scalable 
implementations of universal holonomic quantum computation with qudits. 

\section*{Acknowledgements}
E. S. acknowledges financial support from the Swedish Research Council (VR) through 
Grant No. 2025-05249.


\begin{thebibliography}{99}
\bibitem{zanardi99} P. Zanardi and M. Rasetti,  
Holonomic quantum computation, 
Phys. Lett. A {\bf 264}, 94 (1999). 
\bibitem{sjoqvist12} E. Sj\"{o}qvist, D. M. Tong, L. M. Andersson, B. Hessmo, M. Johansson,
and K. Singh, 
Non-adiabatic holonomic quantum computation, 
New J. Phys. {\bf 14}, 103035 (2012).
\bibitem{wilczek84} F. Wilczek and A. Zee, 
Appearance of gauge structure in simple dynamical systems, 
Phys. Rev. Lett. {\bf 52}, 2111  (1984).
\bibitem{anandan88} J. Anandan, 
Non-adiabatic non-Abelian geometric phase, 
Phys. Lett. A {\bf 133}, 171 (1988). 
\bibitem{duan01} L.-M. Duan, J. I. Cirac, and P. Zoller, 
Geometric manipulation of trapped ions for quantum computation, 
Science {\bf 292}, 1695 (2001). 
\bibitem{wu05} L.-A. Wu, P. Zanardi, and D. A. Lidar, 
Holonomic quantum computation in decoherence-free subspaces, 
Phys. Rev. Lett. {\bf 95}, 130501 (2005). 
\bibitem{xu12} G. F. Xu, J. Zhang, D. M. Tong, E. Sj\"oqvist, and L. C. Kwek, 
Nonadiabatic holonomic quantum computation in decoherence-free subspaces, 
Phys. Rev. Lett. 109, 170501 (2012). 
\bibitem{zhang14} J. Zhang, L.-C. Kwek, E. Sj\"oqvist, D. M. Tong, and P. Zanardi, 
Quantum computation in noiseless subsystems with fast non-Abelian holonomies, 
Phys. Rev. A {\bf 89}, 042302 (2014). 
\bibitem{xu14} G. F. Xu and G. Long, 
Protecting geometric gates by dynamical decoupling, 
Phys. Rev. A {\bf 90}, 022323 (2014). 
\bibitem{zhang15} J. Zhang, T. H. Kyaw, D. M. Tong, E. Sj\"oqvist, and L.-C. Kwek,  
Fast non-Abelian geometric gates via transitionless quantum driving, 
Sci. Rep. {\bf 5}, 18414 (2015). 
\bibitem{xu15} G. F. Xu, C. L. Liu, P. Z. Zhao, and D. M. Tong, 
Nonadiabatic holonomic gates realized by a single-shot implementation, 
Phys. Rev. A {\bf 92}, 052302 (2015).
\bibitem{sjoqvist16} E. Sj\"{o}qvist, 
Nonadiabatic holonomic single-qubit gates in off-resonant $\Lambda$ systems, 
Phys. Lett. A {\bf 380}, 65 (2016).
\bibitem{herterich16} E. Herterich and E. Sj\"oqvist,  
Single-loop multiple-pulse nonadiabatic holonomic quantum gates, 
Phys. Rev. A {\bf 94}, 052310 (2016). 
\bibitem{xue17} Z.-Y. Xue, F.-L. Gu, Z.-P. Hong, Z.-H. Yang, D.-W. Zhang, Y. Hu, and J. Q. You, 
Nonadiabatic holonomic quantum computation with dressed-state qubits, 
Phys. Rev. Applied {\bf 7}, 054022 (2017). 
\bibitem{xu18a} G. F. Xu, D. M. Tong, and E. Sj\"oqvist, 
Path-shortening realizations of nonadiabatic holonomic gates, 
Phys. Rev. A {\bf 98}, 052315 (2018). 
\bibitem{ramberg19} N. Ramberg and E. Sj\"oqvist, 
Environment-assisted holonomic quantum maps, 
Phys. Rev. Lett. {\bf 122}, 140501 (2019).  
\bibitem{liu19} B.-J. Liu, X.-K. Song, Z.-Y. Xue, X. Wang, and M.-H. Yung, 
Plug-and-Play approach to nonadiabatic geometric quantum gates, 
Phys. Rev. Lett. {\bf 123}, 100501 (2019). 
\bibitem{abdumalikov13} A. A. Abdumalikov, J. M. Fink, K. Juliusson, M. Pechal,
S. Berger, A. Wallraff, and S. Filipp, 
Experimental realization of non-Abelian non-adiabatic geometric gates, 
Nature (London) {\bf 496}, 482 (2013).
\bibitem{toyoda13} K. Toyoda, K. Uchida, A. Noguchi, S. Haze, and S. Urabe, 
Realization of holonomic single-qubit operations, 
Phys. Rev. A {\bf 87}, 052307 (2013).
\bibitem{feng13} G. R. Feng, G. F. Xu, and G. L. Long, 
Experimental realization of nonadiabatic holonomic quantum computation, 
Phys. Rev. Lett. {\bf 110}, 190501 (2013).
\bibitem{arroyo14} S. Arroyo-Camejo, A. Lazariev, S. W. Hell, and G. Balasubramanian, 
Room temperature high-fidelity holonomic single-qubit gate on a solid-state spin, 
Nat. Commun. {\bf 5}, 4870 (2014).
\bibitem{zu14} C. Zu, W. B. Wang, L. He, W. G. Zhang, C. Y. Dai, F. Wang, and L. M. Duan, 
Experimental realization of universal geometric quantum gates with solid-state spins, 
Nature (London) {\bf 514}, 72 (2014).
\bibitem{zhou17} B. B. Zhou, P. C. Jerger, V. O. Shkolnikov, F. J. Heremans, 
G. Burkard, and D. D. Awschalom
Holonomic quantum control by coherent optical excitation in diamond, 
Phys. Rev. Lett. {\bf 119}, 140503 (2017). 
\bibitem{sekiguchi17} Y. Sekiguchi, N. Niikura, R. Kuroiwa, H. Kano, and H. Kosaka,  
Optical holonomic single quantum gates with a geometric spin under a zero field, 
Nat. Photonics {\bf 11}, 309 (2017). 
\bibitem{danilin18} S. Danilin, A. Veps\"al\"ainen, and G. S.  Paraoanu, 
Experimental state control by fast non-Abelian holonomic gates with a superconducting qutrit, 
Phys. Scr. {\bf 93},  055101 (2018). 
\bibitem{nagata18} K. Nagata, K. Kuramitani, Y. Sekiguchi, and H. Kosaka, 
Universal holonomic quantum gates over geometric spin qubits with polarised microwaves
Nat. Comm. {\bf 9}, 3227 (2018). 
\bibitem{leroux18} F. Leroux, K. Pandey, R. Rehbi, F. Chevy, C. Miniatura, B. Gr\'emaud, 
and D. Wilkowski, 
Non-Abelian adiabatic geometric transformations in a cold strontium gas, 
Nat. Comm. {\bf 9}, 3580 (2018). 
\bibitem{xu18b} Y. Xu, W. Cai, Y. Ma, X. Mu, L. Hu, Tao Chen, H. Wang, Y. P. Song, 
Z.-Y. Xue, Z.-q. Yin, and L. Sun, 
Single-loop realization of arbitrary nonadiabatic holonomic single-qubit quantum gates 
in a superconducting circuit, 
Phys. Rev. Lett. {\bf 121}, 110501 (2018). 
\bibitem{yan19} T. Yan, B.-J. Liu, K. Xu, C. Song, S. Liu, Z. Zhang, H. Deng, Z. Yan, H. Rong, 
K. Huang, M.-H. Yung, Y. Chen, and D. Yu, 
Experimental realization of nonadiabatic shortcut to non-Abelian geometric gates, 
Phys. Rev. Lett. {\bf 122}, 080501 (2019). 
\bibitem{ai20} M.-Z. Ai, S. Li, Z. Hou, R. He, Z.-H. Qian, Z.-Y. Xue, J.-M. Cui, Y.-F. Huang, 
C.-F. Li, and G.-C. Guo, 
Experimental realization of nonadiabatic holonomic single-qubit quantum gates with 
optimal control in a trapped ion, 
Phys. Rev. Appl. {\bf 14}, 054062 (2020).  
\bibitem{zhao21a} P. Z. Zhao, Z. Dong, Z. Zhang, G. P. Guo, D. M. Tong, and Y. Yin,   
Experimental realization of nonadiabatic geometric gates with a superconducting Xmon qubit, 
Sci. China Phys. Mech. Astron. {\bf 64}, 250362 (2021). 
\bibitem{lu25} C. Lu, X. Wang, and G. Ma,
Experimental realization of special-unitary operations in classical mechanics by nonadiabatic evolutions, 
Phys. Rev. Lett. {\bf 135}, 027201 (2025). 
\bibitem{johansson12} M. Johansson, E. Sj\"oqvist, L. M. Andersson, M. Ericsson, 
B. Hessmo, K. Singh, and D. M. Tong, 
Robustness of nonadiabatic holonomic gates, 
Phys. Rev. A {\bf 86}, 062322 (2012). 
\bibitem{sun16} C. Sun, G. Wang, C. Wu, H. Liu, X.-L. Feng, J.-L. Chen, and K. Xue, 
Non-adiabatic holonomic quantum computation in linear system-bath coupling, 
Sci. Rep. {\bf 6}, 20292 (2016). 
\bibitem{xu17a} G. F. Xu, P. Z. Zhao, T. H. Xing, E. Sj\"oqvist, and D. M. Tong, 
Composite nonadiabatic holonomic quantum computation, 
Phys. Rev. A {\bf 95}, 032311 (2017).
\bibitem{xu17b} G. F. Xu, P. Z. Zhao, D. M. Tong, and E. Sj\"oqvist, 
Robust paths to realize nonadiabatic holonomic gates, 
Phys. Rev. A {\bf 95}, 052349 (2017).
\bibitem{zhao21b} P. Z. Zhao, X. Wu, and D. M. Tong, 
Dynamical-decoupling-protected nonadiabatic holonomic quantum computation, 
Phys. Rev. A {\bf 103}, 012205 (2021).
\bibitem{alves22} G. O. Alves and E. Sj\"oqvist 
Time-optimal holonomic quantum computation, 
Phys. Rev. A {\bf 106}, 032406 (2022).
\bibitem{zhang23} J. Zhang, T. H. Kyaw, S. Filipp, L.-C. Kwek, E. Sj\"oqvist, and D. M. Tong, 
Geometric and holonomic quantum computation, 
Phys. Rep. {\bf 1027}, 1 (2023). 
\bibitem{lanyon09} B. P. Lanyon, M. Barbieri, M. P. Almeida, T. Jennewein, T. C. Ralph, 
K. J. Resch, G. J. Pryde, J. L. O'brien, A. Gilchrist, and A. G. White, 
Simplifying quantum logic using higher-dimensional Hilbert spaces, 
Nat. Phys. {\bf 5}, 134 (2009). 
\bibitem{campbell12} E. T. Campbell, H. Anwar, and D. E. Browne, 
Magic-state distillation in all prime dimensions using quantum Reed-Muller codes, 
Phys. Rev. X {\bf 2}, 041021 (2012).
\bibitem{campbell14} E. T. Campbell, 
Enhanced fault-tolerant quantum computing in $d$-level systems, 
Phys. Rev. Lett. {\bf 113}, 230501 (2014).
\bibitem{andrist15} R. S. Andrist, J. R. Wootton, and H. G. Katzgraber, 
Error thresholds for Abelian quantum double models: Increasing the bit-flip stability of topological 
quantum memory, 
Phys. Rev. A {\bf 91}, 042331 (2015).
\bibitem{michael16}  M. H. Michael, M. Silveri, R. T. Brierley, V. V. Albert, J. Salmilehto, 
L. Jiang and S. M. Girvin, 
New class of quantum error-correcting codes for a bosonic mode, 
Phys. Rev. X {\bf 6}, 031006 (2016).
\bibitem{muralidharan17} S. Muralidharan, C. L. Zou, L. Li, J. Wen, and L. Jiang, 
Overcoming erasure errors with multilevel systems, 
New J. Phys. {\bf 19}, 013026 (2017).
\bibitem{kues17} M. Kues, C. Reimer, P. Roztocki, L. R. Cort\'es, S. Sciara, B. Wetzel, Y. Zhang, 
A. Cino, S. T. Chu, B. E Little, D. J. Moss, L. Caspani, J. Asa\~na, and R. Morandotti, 
On-chip generation of high-dimensional entangled quantum states and their coherent control, 
Nature (London) {\bf 546}, 622 (2017).
\bibitem{moro19} F. Moro, A. J. Fielding, L. Turyanska, and A. Patan\'e, 
Realization of universal quantum gates with spin-qudits in colloidal quantum dots, 
Adv. Quantum Technol. {\bf 2}, 1900017 (2019).
\bibitem{soltamov19} V. A. Soltamov, C. Kasper, A. V. Poshakinskiy, A. N. Anisimov, 
E. N. Mokhov, A. Sperlich, S. A. Tarasenko, P. G. Baranov, G. V. Astakhov, and V. Dyakonov, 
Excitation and coherent control of spin qudit modes in silicon carbide at room temperature, 
Nat. Commun. {\bf 10}, 1678 (2019).
\bibitem{hrmo23} P. Hrmo, B. Wilhelm, L. Gerster, M. W. van Mourik, M. Huber, R. Blatt, 
P. Schindler, T. Monz, and M. Ringbauer, 
Native qudit entanglement in a trapped ion quantum processor, 
Nat. Commun. {\bf 14}, 2242 (2023).  
\bibitem{meng24} Z. Meng, W.-Q. Liu, B.-W. Song, X.-Y. Wang, A.-N. Zhang, and Z.-Q. Yin,  
Experimental realization of high-dimensional quantum gates with ultrahigh fidelity and efficiency, 
Phys. Rev. A {\bf 109}, 022612 (2024). 
\bibitem{rosseaux13} B. Rousseaux, S. Gu\'erin,
and N. V. Vitanov, 
Arbitrary qudit gates by adiabatic passage, 
Phys. Rev. A {\bf 87}, 032328 (2013)
\bibitem{azimi14a} V. Azimi Mousolou, C. M. Canali, and E. Sj\"oqvist,  
Universal non-adiabatic holonomic gates in quantum dots and single-molecule magnets, 
New J. Phys. {\bf 16}, 013029 (2014). 
\bibitem{azimi14b} V. Azimi Mousolou and E. Sj\"oqvist,  
Non-Abelian geometric phases in a system of coupled quantum bits, 
Phys. Rev. A {\bf 89}, 022117 (2014). 
\bibitem{xu21} G. F. Xu, P. Z. Zhao, E. Sj\"oqvist, and D. M. Tong, 
Realizing nonadiabatic holonomic quantum computation beyond the three-level setting, 
Phys. Rev. A {\bf 103}, 052605 (2021).   
\bibitem{andre22} T. Andr\'e and E. Sj\"oqvist, 
Dark path holonomic qudit computation, 
Phys. Rev. A {\bf 106}, 062402 (2022). 
\bibitem{recati02} A. Recati, T. Calarco, P. Zanardi, J. I. Cirac, and P. Zoller, 
Holonomic quantum computation with neutral atoms, 
Phys. Rev. A {\bf 66}, 032309 (2002).  
\bibitem{karle03} R. Karle and J. Pachos, 
Geometrical phases for the G(4,2) Grassmannian manifold, 
J. Math. Phys. {\bf 44}, 2463 (2003). 
\bibitem{shkolnikov20} V. O. Shkolnikov and G. Burkard, 
Effective Hamiltonian theory of the geometric evolution of quantum systems, 
Phys. Rev. A {\bf 101}, 042101 (2020). 
\bibitem{cuadra22} D. Gonz\'alez-Cuadra, T. V. Zache, J. Carrasco, B. Kraus, and P. Zoller, 
Hardware efficient quantum simulation of non-Abelian gauge theories with qudits on Rydberg platforms, 
Phys. Rev. Lett. {\bf 129}, 160501 (2022). 
\bibitem{morris83} J. R. Morris and B. W. Shore, 
Reduction of degenerate two-level excitation to independent two-state systems, 
Phys. Rev. A {\bf 27}, 906 (1983). 
\bibitem{sawicki17} A. Sawicki and K. Karnas, 
Universality of single-qudit gates, 
Ann. Henri Poincar\'e {\bf 18}, 3515 (2017). 
\bibitem{wang20} Y. Wang, Z. Hu, B. C. Sanders, and S. Kais,  
Qudits and high-dimensional quantum computing, 
Front. Phys. {\bf 8}, 589504 (2020). 
\bibitem{ivanov06} P. A. Ivanov, E. S. Kyoseva, and N. V. Vitanov, 
Engineering of arbitrary U$(N)$ transformations by quantum Householder reflections, 
Phys. Rev. A {\bf 74}, 022323 (2006). 
\bibitem{sorensen99} A. S{\o}rensen and K. M{\o}lmer, 
Quantum Computation with ions in thermal motion, 
Phys. Rev. Lett. {\bf 82}, 1971 (1999). 
\end{thebibliography}
\end{document}